\documentclass[showpacs,aps,graphicx,preprintnumbers,amsmath,twocolumn]{revtex4}
\usepackage{mathrsfs}
\usepackage{amsfonts}
\usepackage{amssymb}
\usepackage{graphicx}

\begin{document}

\title{Complete deterministic multi-electron Greenberger-Horne-Zeilinger state analyzer for quantum communication}

\author{Hai-Rui Wei, Bao-Cang Ren, Mei Zhang, Tao Li, and Fu-Guo Deng\footnote{Corresponding author: fgdeng@bnu.edu.cn}}
\address{ Department of Physics, Applied Optics Beijing Area Major
Laboratory, Beijing Normal University, Beijing 100875, China}

\date{\today }

\begin{abstract}
We present a scheme for the multi-electron
Greenberger-Horne-Zeilinger (GHZ) state analyzer, resorting to an
interface between the polarization of a probe photon and the spin of
an electron in a quantum dot embedded in a microcavity. All the
multi-spin GHZ states can be completely discriminated by using
single-photon  detectors and linear optical elements. Our scheme has
some features. First, it is a complete GHZ-state analyzer for
multi-electron spin systems.  Second, the initial entangled states
remain after being identified and they can be used for a successive
task. Third, the electron qubits are static and the photons play a
role of a medium for information transfer, which has a good
application in quantum repeater in which the electron qubits are
used to store the information and the photon qubits are used to
transfer the information between others.
\end{abstract}


\maketitle



\section{Introduction}

Quantum entanglement is a vital resource in quantum information
processing. It enables a powerful quantum computation
\cite{Book,compu1,compu2,longcomputation1,longcomputation2} and it
is also widely used in quantum communication. A maximally entangled
state of three or more particles, called a
Greenberger-Horne-Zeilinger (GHZ) state \cite{r21}, is one of the
most important multiparticle entangled states, and has been
fascinating quantum system to reveal the nonlocality of the quantum
word. It is well known that a complete deterministic Bell-state
analyzer (BSA) is an important prerequisite for many quantum
communication protocols, such as teleportation \cite{telep1,telep2},
entanglement swapping \cite{swapp1,swapp2}, quantum superdense
coding \cite{dense1,dense2,densecoding2}, and some quantum
cryptography protocols
\cite{distri1,distri2,distri3,distri4,distri5,distri6,secr1,secr2,secr3}.
Its extension to a multipartite system, that is, the discrimination
of GHZ states, is useful in quantum computation, quantum
communication, and quantum network. Based on linear optical
elements, a simple GHZ-state analysis scheme, in which one can
distinguish two of eight maximally entangled three-qubit GHZ states,
has been presented in Ref.\cite{GHZPan}. With weak cross-Kerr
nonlinearity, Qian \emph{et al.} \cite{GHZGong1} proposed a
destructive GHZ-state analyzer with a nearly unity probability in
2005.  Based on dipole-induced transparency in a cavity-waveguide
system,  Qian \emph{et al.} \cite{GHZGong2} proposed a
nondestructive GHZ-state analyzer scheme in 2007. A GHZ-state
analyzer scheme based on the input-output relation of a cavity was
presented in Ref.\cite{GHZput} in 2009. Unfortunately, the photons
are absorbed by the detectors and the initial entangled states are
destroyed finally because of destructive measurements. Therefore,
nondestructive analysis of entangled states is also an important and
open problem, and more effective strategies are expected to overcome
the difficulties.

Electron spin in a quantum dot (QD) \cite{QD1,QD2,QD3,QD4,QD5,QD6}
hold great promising in quantum information processing. It has
attracted much attention in recent years. In 2008, Hu \emph{et al.}
\cite{Hu1,Hu2} proposed an interesting device, which is the
singly-electron spin confined in a charged QD inside a microcavity,
and it provides some new methods for quantum information processing.
Based on this spin-QD-cavity unit \cite{Hu1,Hu2}, teleportation
\cite{Hu3}, entanglement swapping \cite{Hu3},  entanglements for
photon-photon \cite{Hu4}, photon-spin and spin-spin \cite{Zhang},
controlled-not gate and phase-shift gates on a hybrid system
\cite{PRL}, and BSA on photon systems \cite{PRL}, entanglement
purification \cite{Wang}, were studied. Moreover, an interesting
quantum repeater scheme \cite{wangtj} based on spin-QD-cavity units
was proposed by Wang, Song, and Long in 2012. In this quantum
repeater scheme, the electron-spin qubits are static and act as the
qubits for storing information. The photon qubits play a role of a
medium for information transfer. When this quantum repeater scheme
is extended to three-party quantum communication or a quantum
network, GHZ state analysis on electron systems is required.
Moreover,  a GHZ state analyzer is essential for long-distance
quantum communication and quantum network, especially in a
multiparty quantum repeater.

In this paper, based on an interface between the polarization of a
probe photon and the spin of an electron in a quantum dot embedded
in a microcavity, we present a scheme for multi-electron GHZ-state
analyzer. All the electron-spin GHZ states can be completely
discriminated by using single-photon detectors and linear optical
elements. Our scheme has some advantages. First, it can be used to
distinguish all the GHZ states, not a part of the states. Second,
the initial entangled electron-spin GHZ states are not changed after
being identified and they can be used for a successive task. Third,
this scheme has a good application in quantum repeater in which the
electron qubits are static and are used to store the information,
and the photon qubits play a role of information transfer. Moreover,
it is essential for quantum communication and quantum network based
on electron-spin qubits.



\section{Physical model}

\begin{figure}[!h]
\begin{center}
\includegraphics[width=7.2 cm,angle=0]{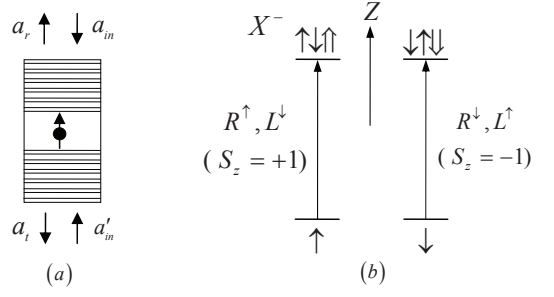}
\caption{(a)A schematic diagram of a singly electron charged QD
confined in an optical resonant microcavity with both mirrors
partially reflective with circular cross-section. (b) A schematic
diagram of the relevant energy levels together with optical spin
selection rules for $X^-$.} \label{figure1}
\end{center}
\end{figure}

We consider a singly electron charged self-assembled GaAs/InAs as
interface quantum dot insider an optical resonant double-side
microcavity with both mirrors partially reflective (see
Fig.\ref{figure1}(a)) \cite{Hu2,PRL}. Singly electron charged QD,
that is, an excess electron is injected into the QD, optical
excitation creates an exciton $X^-$ that consists of two electrons
bound to one hole. $X^-$ dominates the optical property of the
device and the four relevant electronic levels are shown in
Fig.\ref{figure1}(b). In the following, we consider that the dipole
is resonant with the cavity mode and is probed with a resonant
light.  Due to Pauli's exclusion principle, the spin selection rules
for this device is described as follows: (i) If the excess electron
in the spin state $|\uparrow\rangle\equiv|+\frac{1}{2}\rangle$, the
right-circular polarized photon $|R^{\uparrow}\rangle$ with $s_z=+1$
(the superscript uparrow indicates its propagation direction along
the normal direction of the cavity, that is,  $z$ axis) or the
left-circular polarized photon $|L^{\downarrow}\rangle$ with
$s_z=+1$ is couple to the dipole, the photon is reflected by the
cavity, both the polarization and the propagation direction of the
photon will be flipped. However, $|R^{\downarrow}\rangle$  or
$|L^{\uparrow}\rangle$ ($s_z=-1$) is uncouple to the dipole, such
photon transmits through the cavity and acquires a $\pi$ mod $2 \pi$
phase shift relative to a reflected one; (ii) If the excess electron
in the spin state $|\downarrow\rangle\equiv|-\frac{1}{2}\rangle$, in
the same way, $|R^{\downarrow}\rangle$  or $|L^{\uparrow}\rangle$
($s_z=-1$) are are reflected by the cavity, $|R^{\uparrow}\rangle$
and $|L^{\downarrow}\rangle$ ($s_z=+1$) are transmitted through the
cavity. Therefore, the rule of the input states changed under the
interaction between the photons with $s_z =\pm1$ and the cavity is
described as follows:
\begin{eqnarray}                            \label{eq.1}
|R^\uparrow,\uparrow\rangle&\rightarrow&|L^\downarrow,\uparrow\rangle,\;\;\;\;\;\;\;\;\;
|L^\uparrow,\uparrow\rangle\rightarrow-|L^\uparrow,\uparrow\rangle, \nonumber\\
|R^\downarrow,\uparrow\rangle&\rightarrow&-|R^\downarrow,\uparrow\rangle,\;\;\;\;\;\;
|L^\downarrow,\uparrow\rangle\rightarrow|R^\uparrow,\uparrow\rangle,\nonumber\\
|R^\uparrow,\downarrow\rangle&\rightarrow&-|R^\uparrow,\uparrow\rangle,\;\;\;\;\;\;
|L^\uparrow,\downarrow\rangle\rightarrow|R^\downarrow,\downarrow\rangle,\nonumber\\
|R^\downarrow,\downarrow\rangle&\rightarrow&|L^\uparrow,\uparrow\rangle,\;\;\;\;\;\;\;\;\;
|L^\downarrow,\downarrow\rangle\rightarrow-|L^\downarrow,\downarrow\rangle.
\end{eqnarray}
In Fig.\ref{figure1}, $|\Uparrow\rangle$ and $|\Downarrow\rangle$
represent the hole-spin states $|+\frac{3}{2}\rangle$ and
$|-\frac{3}{2}\rangle$, respectively.

\section{Complete three-electron-spin GHZ-state analyzer}

\begin{figure}[!h]
\begin{center}
\includegraphics[width=8.0 cm,angle=0]{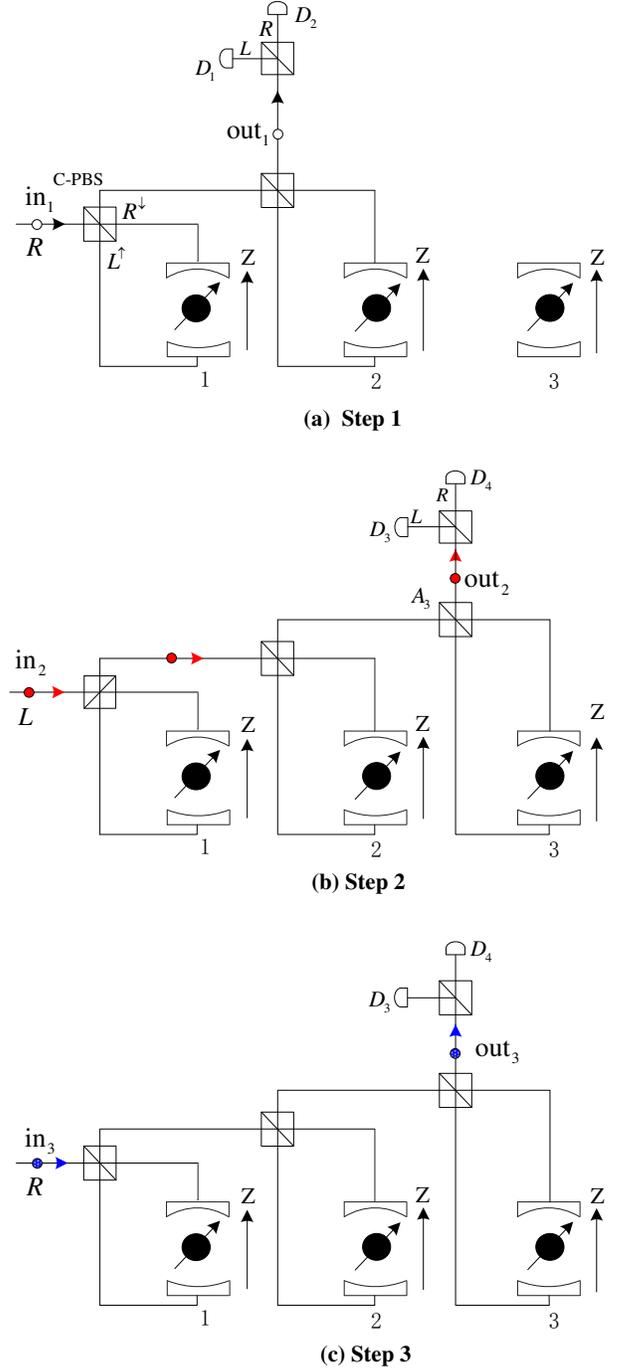}
\caption{(Color online) The schematic diagram for distinguishing the
eight three-spin GHZ states. C-PBS is a polarizing beam splitter in
the circular basis, and it transmits a right-circular polarization
photon $|R\rangle$ and reflects a left-circular polarization photon
$|L\rangle$. $D_i$ ({$i=1,2,3,4$}) are four photon detectors.}
\label{figure2}
\end{center}
\end{figure}

Before starting to discuss our GHZ-state analyzer, let us first
extend the study of creating the two-qubit Bell states in
Ref.\cite{Hu1} to the three-spin GHZ states via QD-spin-cavity units
discussed above. The principle of our  scheme for GHZ-state analysis
is shown in Fig.\ref{figure2}(a) and Fig.\ref{figure2}(b). Consider
three uncorrelated excess spins are in  arbitrary states
$|\psi_i^s\rangle=\alpha_i|\uparrow\rangle+\beta_i|\downarrow\rangle$
($i=1, 2, 3$) and the probe photon 1 and the  probe photon 2 are in
the polarization state $|R_1^\downarrow\rangle$ and
$|L_2^\uparrow\rangle$, respectively. The two photons are sent
through the input port and the polarizing beam splitter in the
circular basis (C-PBS) one after another. The first C-PBS is rotated
by $90^\circ$ after the first photon $|R_1^\downarrow\rangle$ (label
$in_1$) passes through it, so that the second photon
$|L_2^\uparrow\rangle$ (label $in_2$) deserts the first cavity and
injects directly into the cavities 2 and 3 in sequence, and emits
from the output port $out_2$. The  first photon
$|R_1^\downarrow\rangle$ (label $in_1$) passes through the cavities
1 and 2 in sequence, and emits from the  port $out_1$. Based on the
rules in Eq.(\ref{eq.1}), the evolution of the system  composed of
two photons and  three electron spins can be written as
\begin{eqnarray}                            \label{eq.2}
&&|R_1^\downarrow\rangle \otimes|L_2^\uparrow\rangle
\otimes|\psi_1^{s}\rangle \otimes|\psi_2^{s}\rangle \otimes
|\psi_3^{s}\rangle\nonumber\\&\rightarrow&\;\;\;
|R_1^\downarrow\rangle|L_2^\uparrow\rangle(\alpha_1\alpha_2\alpha_3|\uparrow_1\uparrow_2\uparrow_3\rangle
+\beta_1\beta_2\beta_3|\downarrow_1\downarrow_2\downarrow_3\rangle \nonumber\\
&&-|R_1^\downarrow\rangle|R_2^\downarrow\rangle(\alpha_1\alpha_2\beta_3|\uparrow_1\uparrow_2\downarrow_3\rangle
+\beta_1\beta_2\alpha_3|\downarrow_1\downarrow_2\uparrow_3\rangle \nonumber\\
&&+|L_1^\uparrow\rangle|R_2^\downarrow\rangle(\alpha_1\beta_2\alpha_3|\uparrow_1\downarrow_2\uparrow_3\rangle
+\beta_1\alpha_2\beta_3|\downarrow_1\uparrow_2\downarrow_3\rangle \nonumber\\
&&-|L_1^\uparrow\rangle|L_2^\uparrow\rangle(\alpha_1\beta_2\beta_3|\uparrow_1\downarrow_2\downarrow_3\rangle
+\beta_1\alpha_2\alpha_3|\downarrow_1\uparrow_2\uparrow_3\rangle).
\end{eqnarray}
When the two-photon system is in the states
$|R_1\rangle|L_2\rangle$, $|R_1\rangle|R_2\rangle$,
$|L_1\rangle|R_2\rangle$ and $|L_1\rangle|L_2\rangle$, the
three-spin system is in the states $|\psi_0\rangle$,
$|\psi_1\rangle$, $|\psi_2\rangle$ and $|\psi_3\rangle$,
respectively. Here
\begin{eqnarray}                                                \label{eq.3}
|\psi_0\rangle &=&
2(\alpha_1\alpha_2\alpha_3|\uparrow_1\uparrow_2\uparrow_3\rangle+\beta_1\beta_2\beta_3|\downarrow_1\downarrow_2\downarrow_3\rangle),\nonumber\\
|\psi_1\rangle &=&
2(\alpha_1\alpha_2\beta_3|\uparrow_1\uparrow_2\downarrow_3\rangle+\beta_1\beta_2\alpha_3|\downarrow_1\downarrow_2\uparrow_3\rangle),\nonumber\\
|\psi_2\rangle &=&
2(\alpha_1\beta_2\alpha_3|\uparrow_1\downarrow_2\uparrow_3\rangle+\beta_1\alpha_2\beta_3|\downarrow_1\uparrow_2\downarrow_3\rangle),\nonumber\\
|\psi_3\rangle &=&
2(\alpha_1\beta_2\beta_3|\uparrow_1\downarrow_2\downarrow_3\rangle+\beta_1\alpha_2\alpha_3|\downarrow_1\uparrow_2\uparrow_3\rangle).
\end{eqnarray}
By setting the coefficient $\alpha_i$ and $\beta_i$ to
$\frac{1}{\sqrt2}$, or any one of $\alpha_i$ and $\beta_i$ to
$-\frac{1}{\sqrt2}$ and the others to $\frac{1}{\sqrt2}$, one can
obtain the eight maximally entangled three-spin GHZ states
\begin{eqnarray}                                                \label{eq.4}
|\psi^{\pm}_0\rangle &=& \frac{1}{\sqrt2}(|\uparrow_1\uparrow_2\uparrow_3\rangle \pm |\downarrow_1\downarrow_2\downarrow_3\rangle), \nonumber\\
|\psi^{\pm}_1\rangle &=& \frac{1}{\sqrt2}(|\uparrow_1\uparrow_2\downarrow_3\rangle \pm  |\downarrow_1\downarrow_2\uparrow_3\rangle), \nonumber\\
|\psi^{\pm}_2\rangle &=& \frac{1}{\sqrt2}(|\uparrow_1\downarrow_2\uparrow_3\rangle \pm |\downarrow_1\uparrow_2\downarrow_3\rangle),\nonumber\\
|\psi^{\pm}_3\rangle&=&\frac{1}{\sqrt2}(|\uparrow_1\downarrow_2\downarrow_3\rangle\pm|\downarrow_1\uparrow_2\uparrow_3\rangle).
\end{eqnarray}

Now, we explain how to implement a complete deterministic three-spin
GHZ-state analyzer. The setup of our scheme is depicted in
Fig.\ref{figure2}. It consists of three parts: (i) the photon 1 in
the polarization state $|R^\downarrow_1\rangle$ (label $in_1$) is
injected into the cavity 1 and the cavity 2 in succession and it is
detected at the port $out_1$; (ii) the photon 2 in the polarization
state $|L_2^\uparrow\rangle$ (label $in_2$) is injected into the
cavity 2 and cavity 3 in sequence and a $90^\circ$ rotation on the
first C-PBS is performed before the second photon
$|L_2^\uparrow\rangle$ passes through it (so that the photon
$|L_2^\uparrow\rangle$ deserts the the first cavity and injects
directly into the cavity 2 and the cavity 3 in sequence), and it is
detected at the port  $out_2$; (iii) a Hadamard transformation
(e.g., using a $\pi/2$ microwave pulse)
$|\uparrow\rangle\rightarrow\frac{1}{\sqrt2}(|\uparrow\rangle+|\downarrow\rangle)$,
$|\downarrow\rangle\rightarrow\frac{1}{\sqrt2}(|\uparrow\rangle-|\downarrow\rangle)$
on each spin is applied and then the probe photon 3 in the
polarization state $|R_3^\downarrow\rangle$ (label $in_3$) is
injected into the cavity 1, the cavity 2 and the cavity 3 in
sequence with a $90^\circ$ rotation on the first C-PBS again before
the photon 3 passes through it.

In the first step, one can distinguish $|\psi^{\pm}_0\rangle$ and
$|\psi^{\pm}_1\rangle$ from $|\psi^{\pm}_2\rangle$ and
$|\psi^{\pm}_3\rangle$ by simply discriminating the instance which
detector is clicked by the photon. After the photon passes through
cavity 1 and cavity 2 in succession, it will emit from the port
$out_1$ and is in the state
\begin{eqnarray}                                                \label{eq.5}
|R^\downarrow\rangle|\psi^{\pm}_0\rangle &\xrightarrow{\text{through cavity 1, 2}}& |R^\downarrow\rangle|\psi^{\pm}_0\rangle, \nonumber\\
|R^\downarrow\rangle|\psi^{\pm}_1\rangle &\xrightarrow{\text{through cavity 1, 2}}& |R^\downarrow\rangle|\psi^{\pm}_1\rangle, \nonumber\\
|R^\downarrow\rangle|\psi^{\pm}_2\rangle &\xrightarrow{\text{through cavity 1, 2}}& -|L^\uparrow\rangle|\psi^{\pm}_2\rangle,\nonumber\\
|R^\downarrow\rangle|\psi^{\pm}_3\rangle&\xrightarrow{\text{through
cavity 1, 2,}}&-|L^\uparrow\rangle|\psi^{\pm}_3\rangle.
\end{eqnarray}
Hence, we can distinguish $|\psi^{\pm}_0\rangle$ and
$|\psi^{\pm}_1\rangle$ (corresponds to the instance that the photon
is detected by detector $D_2$) from $|\psi^{\pm}_2\rangle$ and
$|\psi^{\pm}_3\rangle$ (corresponds to the instance that the photon
is detected by detector $D_1$).

As the same  as that in the first step, after the photon  2 passes
through the cavity 2 and the cavity 3 in succession, the evolution
of the system can be written as
\begin{eqnarray}                                                \label{eq.6}
|L^\uparrow\rangle|\psi^{\pm}_0\rangle &\xrightarrow{\text{through cavity 2, 3}}& |L^\uparrow\rangle|\psi^{\pm}_0\rangle, \nonumber\\
|L^\uparrow\rangle|\psi^{\pm}_1\rangle &\xrightarrow{\text{through cavity 2, 3}}& -|R^\downarrow\rangle|\psi^{\pm}_1\rangle, \nonumber\\
|L^\uparrow\rangle|\psi^{\pm}_2\rangle &\xrightarrow{\text{through cavity 2, 3}}& -|R^\downarrow\rangle|\psi^{\pm}_2\rangle,\nonumber\\
|L^\uparrow\rangle|\psi^{\pm}_3\rangle&\xrightarrow{\text{through
cavity 2, 3}}&|L^\uparrow\rangle|\psi^{\pm}_3\rangle.
\end{eqnarray}
That is, one can distinguish $|\psi^{\pm}_0\rangle$  (corresponds to
the photon is detected by detector $D_3$) from
$|\psi^{\pm}_1\rangle$ (corresponds to the detector $D_4$) and
distinguish $|\psi^{\pm}_2\rangle$ (corresponds to the detector
$D_4$) from $|\psi^{\pm}_3\rangle$ (corresponds to the detector
$D_3$).

With two steps of the quantum nondemolition detectors (QND) shown in
Fig.\ref{figure2}, the eight GHZ states are divided into four
groups, that is, $|\psi^{\pm}_0\rangle $, $|\psi^{\pm}_1\rangle$,
$|\psi^{\pm}_2\rangle$ and $|\psi^{\pm}_3\rangle$. The next task is
only to distinguish the different relative phases in each group and
it can be accomplished by applying Hadamard transformation on each
spin as the eight GHZ states will be transformed into
\begin{eqnarray}                                                \label{eq.7}
|\psi^{+}_0\rangle &\xrightarrow{\text{$H$}}&
\frac{1}{2}(|\uparrow_1\uparrow_2\uparrow_3\rangle
+|\uparrow_1\downarrow_2\downarrow_3\rangle+|\downarrow_1\uparrow_2\downarrow_3\rangle+|\downarrow_1\downarrow_2\uparrow_3\rangle), \nonumber\\
|\psi^{-}_0\rangle &\xrightarrow{\text{$H$}}&
\frac{1}{2}(|\uparrow_1\uparrow_2\downarrow_3\rangle
+|\uparrow_1\downarrow_2\uparrow_3\rangle+|\downarrow_1\uparrow_2\uparrow_3\rangle+|\downarrow_1\downarrow_2\downarrow_3\rangle), \nonumber\\
|\psi^{+}_1\rangle &\xrightarrow{\text{$H$}}&
\frac{1}{2}(|\uparrow_1\uparrow_2\uparrow_3\rangle
+|\downarrow_1\downarrow_2\uparrow_3\rangle-|\uparrow_1\downarrow_2\downarrow_2\rangle-|\downarrow_1\uparrow_2\downarrow_3\rangle),\nonumber\\
|\psi^{-}_1\rangle&\xrightarrow{\text{$H$}}&\frac{1}{2}(|\uparrow_1\downarrow_2\uparrow_3\rangle
+|\downarrow_1\uparrow_2\uparrow_3\rangle-|\uparrow_1\uparrow_2\downarrow_3\rangle-|\downarrow_1\downarrow_2\downarrow_3\rangle),\nonumber\\
|\psi^{+}_2\rangle &\xrightarrow{\text{$H$}}&
\frac{1}{2}(|\uparrow_1\uparrow_2\uparrow_3\rangle
-|\downarrow_1\downarrow_2\uparrow_3\rangle-|\uparrow_1\downarrow_2\downarrow_3\rangle+|\downarrow_1\uparrow_2\downarrow_3\rangle), \nonumber\\
|\psi^{-}_2\rangle &\xrightarrow{\text{$H$}}& \frac{1}{2}(|\downarrow_1\uparrow_2\uparrow_3\rangle-|\uparrow_1\downarrow_2\uparrow_3\rangle
+|\uparrow_1\uparrow_2\downarrow_3\rangle-|\downarrow_1\downarrow_2\downarrow_3\rangle), \nonumber\\
|\psi^{+}_3\rangle &\xrightarrow{\text{$H$}}&
\frac{1}{2}(|\uparrow_1\uparrow_2\uparrow_3\rangle
-|\downarrow_1\downarrow_2\uparrow_3\rangle+|\uparrow_1\downarrow_2\downarrow_3\rangle-|\downarrow_1\uparrow_2\downarrow_3\rangle),\nonumber\\
|\psi^{-}_3\rangle&\xrightarrow{\text{$H$}}&\frac{1}{2}(|\downarrow_1\uparrow_2\uparrow_3\rangle-|\uparrow_1\downarrow_2\uparrow_3\rangle
-|\uparrow_1\uparrow_2\downarrow_3\rangle+|\downarrow_1\downarrow_2\downarrow_3\rangle).\nonumber\\
\end{eqnarray}
Obviously, one can distinguish $|\psi^{+}_i\rangle $ (corresponds to
the instance that the number of the state $|\uparrow\rangle$ is odd,
that is, corresponds to the detector $D_4$) from
$|\psi^{-}_i\rangle$ (corresponds to an even number of the state
$|\uparrow\rangle$, that is, corresponds to the detector $D_3$).

From the  analysis above, one can see that the role of the photon is
only the medium for distinguishing the eight GHZ electron-spin
states and the spins are not destroyed. With the first step, one can
distinguish $|\psi^{\pm}_0\rangle$, $|\psi^{\pm}_1\rangle$ from
$|\psi^{\pm}_2\rangle$, $|\psi^{\pm}_3\rangle$. The second QND can
distinguish $|\psi^{\pm}_0\rangle$ from $|\psi^{\pm}_1\rangle$,
$|\psi^{\pm}_2\rangle$ from $|\psi^{\pm}_3\rangle$, and then all of
the GHZ states are divided into four groups $|\psi^{\pm}_i\rangle$.
With the third step,  the two states with different relative phases
are distinguished. The possible measurement results for each
three-spin GHZ state in each step is given in Table \ref{table1}.

\begin{table}
\caption {Output results for each step of complete three-spin GHZ
state analyzer.\\}

\begin{ruledtabular}
\begin{tabular}{cccc}
& \multicolumn {3}{c}{Results} \\
    State                  &          Step 1          &            Step 2               &        Step 3           \\
   \hline

  $|\psi^{+}_0\rangle$     &  $|R_1^\downarrow\rangle$  &  $|L_2^\uparrow\rangle$       &   odd $|\uparrow\rangle$ $(|R_3^\downarrow\rangle)$  \\

  $|\psi^{-}_0\rangle$     &  $|R_1^\downarrow\rangle$  &  $|L_2^\uparrow\rangle$       &   even $|\uparrow\rangle$  $(|L_3^\uparrow\rangle)$  \\

  $|\psi^{+}_1\rangle$    &  $|R_1^\downarrow\rangle$   &  $|R_2^\downarrow\rangle$     &   odd $|\uparrow\rangle$ $(|R_3^\downarrow\rangle)$   \\

  $|\psi^{-}_1\rangle$    &  $|R_1^\downarrow\rangle$  &   $|R_2^\downarrow\rangle$     &   even $|\uparrow\rangle$ $(|L_3^\uparrow\rangle)$\\

  $|\psi^{+}_2\rangle$     &  $|L_1^\uparrow\rangle$   &   $|R_2^\downarrow\rangle$     &   odd $|\uparrow\rangle$  $(|R_3^\downarrow\rangle)$  \\

  $|\psi^{-}_2\rangle$     &  $|L_1^\uparrow\rangle$    &  $|R_2^\downarrow\rangle$     &   even $|\uparrow\rangle$  $(|L_3^\uparrow\rangle)$   \\

  $|\psi^{+}_3\rangle$    &   $|L_1^\uparrow\rangle$   &  $|L_2^\uparrow\rangle$        &   odd $|\uparrow\rangle$  $(|R_3^\downarrow\rangle)$   \\

  $|\psi^{-}_3\rangle$    &   $|L_1^\uparrow\rangle$    &  $|L_2^\uparrow\rangle$       &   even $|\uparrow\rangle$  $(|L_3^\uparrow\rangle)$ \\
\end{tabular} \label{table1}
\end{ruledtabular}
\end{table}

\section{Complete multi-electron-spin GHZ-state analyzer}

Combing Eq.(\ref{eq.1}) and Fig.\ref{figure2}, we find that the
polarization of the probe photon will remain after the probe photon
$|R^\downarrow\rangle$ or $|L^\uparrow\rangle$ interacts with the
two cavities,  if the two excess electron spins in the cavities are
parallel ($|\uparrow_1,\uparrow_2,\rangle$ or
$|\downarrow_1,\downarrow_2,\rangle$); otherwise, the polarization
of the probe photon will be flipped. That is, the present scheme for
three-electron GHZ-state analyzer (see Fig.\ref{figure2}) can be
generalized to multi-electron-spin qubits. The $N$-qubit GHZ states
can be written as
\begin{eqnarray}                                                \label{eq.8}
|\Psi^\pm_i\rangle&=&\frac{|B_N(i)\rangle\pm|B_N(2^N-1-i)\rangle}{\sqrt2},\nonumber\\
&=&\frac{|\uparrow_1,i_2,i_3,\ldots,i_N\rangle\pm
|\downarrow_1,\overline{i}_2,\overline{i}_3,\ldots,\overline{i}_N\rangle}{\sqrt2},
\end{eqnarray}
where $i_k=\uparrow, \downarrow$ ($\forall k=2,3,\ldots, N$) and
$\overline{i}_k$ is the result of flipping each bit of $i_k$, and
$B_N(i)=\uparrow_1,i_2,i_3,\ldots,i_N$ is the binary notation and
$i=0, 1, \ldots, 2^{N-1}-1$.

\begin{figure*}[htb]        
\begin{center}
\includegraphics[width=14 cm,angle=0]{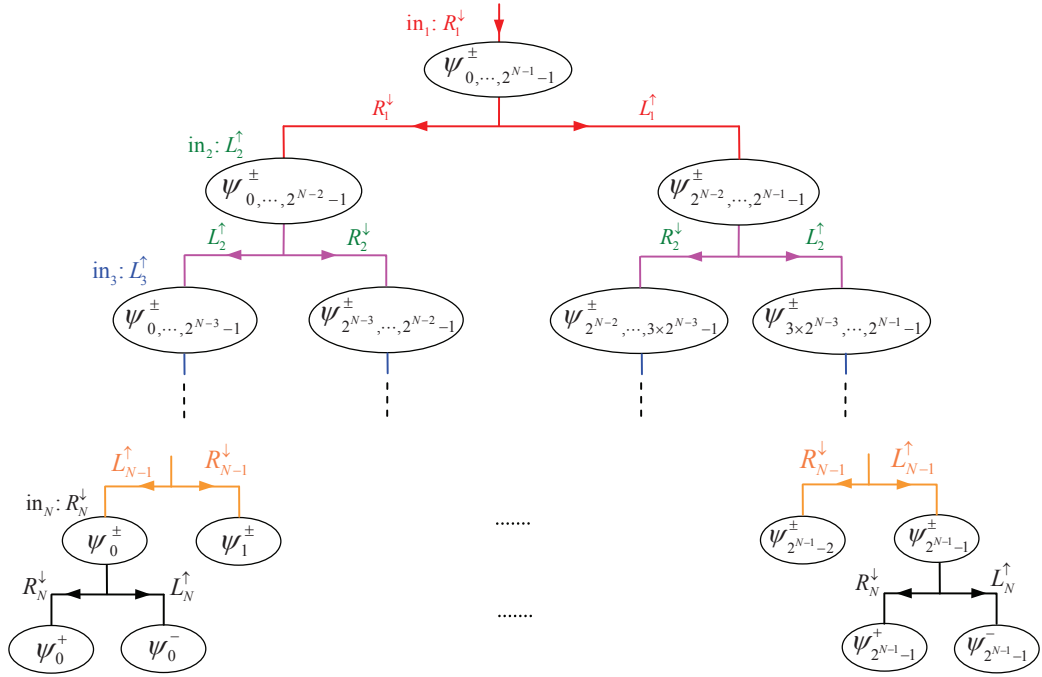}
\caption{(Color online) The tree diagram for our complete
deterministic multi-electron-spin GHZ-state analyzer. With the first
step, one can distinguish $|\Psi^\pm_{0,\ldots,2^{N-2}-1}\rangle$
(corresponds to the instance that $|R_1^\downarrow\rangle$ is
detected) from $|\Psi^\pm_{2^{N-2},\ldots,2^{N-1}-1}\rangle$
(corresponds to the instance that $|L_1^\uparrow\rangle$ is
detected); With the second step, take the second probe photon
$|L_2^\uparrow\rangle$ (label $in_2$) as an example, one can
distinguish $|\Psi^\pm_{0,\ldots,2^{N-3}-1}\rangle$ (corresponds to
the instance that $|L_2^\uparrow\rangle$ is detected) from
$|\Psi^\pm_{2^{N-3},\ldots,2^{N-2}-1}\rangle$ (corresponds to the
instance that $|R_2^\downarrow\rangle$ is detected) and distinguish
$|\Psi^\pm_{3\times2^{N-3},\ldots,2^{N-1}-1}\rangle$ (corresponds to
the instance that $|L_2^\uparrow\rangle$ is detected) from
$|\Psi^\pm_{2^{N-2}, \dots, 3\times2^{N-3}-1}\rangle$ (corresponds
to the instance that
 $|R_2^\downarrow\rangle$ is detected); With the last
step, one can distinguish $|\Psi_i^+\rangle$ (corresponds to the
instance that $|R_N^\downarrow\rangle$ is detected) from
$|\Psi_i^-\rangle$ (corresponds to the instance that
$|L_N^\uparrow\rangle$ is detected).} \label{figure3}
\end{center}
\end{figure*}

The complete deterministic multi-electron GHZ-state analysis can
also be accomplished with three steps: (i) the first probe photon in
state $|R_1^\downarrow\rangle$ is injected only into the cavity 1
and the cavity 2 in succession and it is detected; (ii) the $j_{th}$
probe photon in the state $|L_j^\uparrow\rangle$ deserts the cavity
1, $\ldots$, the cavity $j-1$  and is injected directly into the
cavity $j$ and the cavity $j+1$ and it is detected ($j=2$, \ldots
,$N-1$);
 (iii) a Hadamard transformation on each spin is applied, and  $N_{th}$ probe photon in state
$|R_N^\downarrow\rangle$ is injected into all the cavities in
succession and it is detected.  The possible measurement results
corresponds to its probe photon is given in Fig.\ref{figure3}.

\section{Discussion and summary}

We have proposed  a scheme for three-electron-spin GHZ-state
analyzer and extend it to multi-electron-spin case. The three-spin
(multi-spin) GHZ state is prepared in QD spins in optical resonant
microcavities. A QD spins can be used to store and process quantum
information due to the long electron-spin coherence time ($\sim\mu
s$) \cite{us1,us2}, which is limited by the spin-relaxation time
($\sim ms$) \cite{ms1,ms2}. Spin manipulation is well developed
using pulsed magnetic-resonance techniques, and single spin
detection can be implemented by distinguishing whether the
polarization of the probe photon is flipped or not after it passes
through the cavity. Assume that the probe photon is in the state
$|R^\downarrow\rangle$, the flip of the polarization indicates the
fact that the electron is in the state $|\downarrow\rangle$;
otherwise, the electron is in the state $|\uparrow\rangle$.

Our scheme can distinguish all GHZ states with single-photon
detectors and linear-optical elements. In our scheme, the initial
entangled states remain after being identified and they can be used
for a successive task. Moreover, the electron-spin qubits are
static, and the photons are only a medium for information transfer.
The multi-particle GHZ-state analyzer is essential for
multi-particle generalizations of quantum teleportation, quantum
dense coding, entanglement swapping, and quantum networks. The
present GHZ analysis scheme may be useful in quantum computing,
quantum communication and quantum network.

\section*{ACKNOWLEDGEMENTS}

This work is supported by the National Natural Science Foundation of
China under Grant Nos. 10974020 and 11174039,  NCET-11-0031, and the
Fundamental Research Funds for the Central Universities.

\end{document}